%% file: main.tex
\documentclass[twocolumn,tighten]{aastex63}

\usepackage{times,natbib,graphicx,amsmath,multirow,xspace}
\usepackage{xcolor}
\usepackage{lineno}
\let\tablenum\relax

\usepackage{siunitx}
\usepackage{hyperref}

\newcommand{\nustar}{NuSTAR\xspace}

\newcommand{\chandra}{Chandra\xspace}
\newcommand{\nicer}{NICER\xspace}

\newcommand{\ms}{$M_{\odot}$\xspace}

\newcommand{\rin}{$R_{\rm in}$\xspace}
\newcommand{\rg}{$R_{g}$\xspace}
\newcommand{\risco}{$R_{\mathrm{ISCO}}$\xspace}
\newcommand{\cps}{counts s$^{-1}$\xspace}

\newcommand{\relxillns}{{\sc relxillNS}\xspace}

\newcommand{\reflionx}{{\sc reflionx}\xspace}

\newcommand{\xrism}{XRISM\xspace}
\newcommand{\ixpe}{IXPE\xspace}
\newcommand{\xstar}{{\sc xstar}\xspace}

\shorttitle{XRISM-NuSTAR View of Serpens X-1}
\shortauthors{Hall et al.}

\begin{document}

\title{Resolving structure within the iron line profile of Serpens X-1 with XRISM and NuSTAR}

\correspondingauthor{H. Hall}
\email{hk5269@wayne.edu}

\author[0009-0000-4409-7914]{H. Hall}
\affiliation{Department of Physics \& Astronomy, Wayne State University, 666 West Hancock Street, Detroit, MI 48201, USA}
\author[0000-0002-8961-939X]{R.~M.~Ludlam}
\affiliation{Department of Physics \& Astronomy, Wayne State University, 666 West Hancock Street, Detroit, MI 48201, USA}
\author[0000-0002-8294-9281]{E.Cackett}
\affiliation{Department of Physics \& Astronomy, Wayne State University, 666 West Hancock Street, Detroit, MI 48201, USA}
\author[0000-0003-2869-7682]{J. M. Miller}
\affiliation{Department of Astronomy, University of Michigan, 1085 South University Ave, Ann Arbor, MI 48109-1107, USA}
\author[0000-0003-3828-2448]{J. A. Garc\'ia}
\affiliation{NASA Goddard Space Flight Center, Greenbelt, MD 20771, USA}
\affiliation{Cahill Center for Astronomy and Astrophysics, California Institute of Technology, 1200 E. California Blvd, MC 290-17, Pasadena, CA, 91125, USA}
\begin{abstract}
We present the first simultaneous \xrism and \nustar observations of the neutron star low mass X-ray binary Serpens X-1. We perform spectral modeling on the joint observations of \xrism/Resolve, \xrism/Xtend, and \nustar, testing Comptonization and double thermal continuum model prescriptions. We find that a hybrid double thermal model adequately describes the underlying continuum of the source while the Comptonization models predict largely unphysical parameter values. We perform reflection modeling and confirm with tight constraints an inclination of $5^{\circ}\pm1^{\circ}$ and inner disk radius of $6.6\pm0.6$ \rg for the source. We show that the spectral resolution of \xrism/Resolve allows for the determination of a unique inner disk radius, which has the potential to constrain neutron star spin in other systems. We discuss discrepancies in spectral shape between \xrism/Resolve and \nustar above $\sim8$ keV. We also present lightcurve and spectral analyses of 8 Type-I X-ray bursts that occurred during the observations.    
\end{abstract}


\section{Introduction} \label{sec:intro}
Neutron star (NS) low mass X-ray binaries (LMXB) consist of a NS and a $\lesssim 1$ \ms companion that transfers material via a disk onto the NS through Roche-lobe overflow. This accretion can be persistent, i.e., relatively constant with time, or transient, cycling between periods of outburst and quiescence. Bright, persistent NS LXMBs can be further divided into two categories based on their variability and X-ray spectral properties. The classes, ``atoll" and ``Z", are so named due to the shapes traced out in their hardness-intensity (HID) and color-color  diagrams \citep{1989A&A...225...79H}. Atolls are less luminous than Z sources, typically exhibiting luminosities in the range of $\sim$ 0.01-0.5 $L_{\rm Edd}$ and can be either found in the so-called soft ``banana" branch or hard ``island" state \citep{Homan_2010}. When found in the harder state, the spectrum is dominated by Comptonized emission from the coronal region with an additional thermal component with a temperature $\lesssim$ 1 keV \citep{2000_Barret,2001_Church_Balucinska}. The spectrum seen in the softer state is dominated by thermal emission with a weaker Comptonized component. There are several models that have been used to describe these systems in the soft state. The Western model \citep{1988_White_west} is comprised of a single temperature blackbody and Comptonized disk emission while the Eastern model \citep{1989_Mitsuda_east} features a multi-colored blackbody for the disk along with a thermally Comptonized emission  component which arises due to the inverse-Compton scattering of photons off relativistic electrons in the corona. Also, a hybrid double thermal model has been used which includes both a single and multi-temperature blackbody component for soft emission in conjunction with weakly Comptonized higher energy emission in the form of a power-law \citep{2007_Lin}.

The accretion disk surrounding the NS can be externally illuminated from either the NS surface or a boundary/spreading layer (BL/SL) \citep{Popham_2001,Inogamov_Sunyaev_1999}, or from the non-thermal emission from the corona \citep{1991SvAL...17..409S}. These photons are absorbed, reprocessed, and re-emitted by the disk in the form of a `reflected' blackbody spectrum with superimposed atomic features (both emission and absorption). This component is known as the reflection spectrum of X-ray binaries \citep{2005_Ross_Fabian,2010_Garcia,Garc_a_2022}. The atomic features in the reflection spectrum are broadened asymmetrically from a combination of Doppler, special and general realtivistic effects arising in the inner regions of the accretion disk \citep{1989_Fabian,2013_Dauser}. The strength of gravitational redshift becomes amplified the closer the emitting material is to the compact object \citep{2000_Fabian_Broad_Fe_AGN}. The binary inclination impacts the strength of the Doppler effects, with higher inclinations leading to more severe broadening \citep{2010_Dauser}. Thus, examining this reflection component can reveal key information on the central NS (i.e., size and magnetic field strength) as well as the physics taking place within the accretion disk \citep{2010_Cackett,Miller_2013,2017_Ludlam_a,2017_Ludlam_b,Garc_a_2022,2024_Ding}. Radius measurements of NS are critical to uncovering the pressure-energy density relation, otherwise known as the equation of state (EOS), of matter within the NS (see \citealt{2021_Lattimer} for review). Placing better constraints on NS mass and radii is paramount in its determination. Reflection models, like \reflionx \citep{2005_Ross_Fabian} and \relxillns \citep{Garc_a_2022} have been used to accurately describe the full reflected spectrum produced in these systems and give robust estimates of system parameters.

NS LMXBs also show unstable thermonuclear bursts, known as Type-I X-ray bursts (see \citealt{1993_Lewin,2006_Strohmayer_Bild,2006_Schatz,Galloway_2021} for reviews). They arise when the accreted material (typically hydrogen and helium) reaches sufficient pressure and temperature on the NS surface to ignite in an unstable runaway process which consumes most of the accreted fuel. The nature of the bursts depends primarily on mass-accretion rate \citep{1981_Fujimoto,1987_Fujimoto}, however, factors like metallicity of the accreted material and compactness of the NS can also have an effect \citep{1998_Bildsten,2003_Narayan,2013_Degenaar}. They are evident in the light curves as a sharp rapid increase in intensity followed by a slow decay, often modeled with a fast-rise-exponential-decay function (Eq.~\ref{eq:FRED}).

Their spectra is typically best represented by a variable Planck distribution with blackbody temperatures ranging from 2--3 keV during the burst and around 1 keV in the tail \citep{2013_Degenaar,Pike:2021uzn}. The spectrum is largely devoid of the spectral features seen from reflection. In theory, the spectrum can show distinct absorption features if the photosphere is not fully ionized. Owing to the fact that these absorption features arise close to the surface of the NS, a determination of their redshift could give a measure of NS compactness (the ratio of NS mass-radius), a key metric in determining the NS EOS \citep{2006_Ozel,2013_Degenaar,2018_Yoneda,Galloway_2021}. 

Serpens X-1 (hereafter Ser X-1) is a bright persistently accreting atoll NS LMXB located at a distance of $7.7\pm0.9$ kpc away \citep{Galloway_2008}. The source has shown some variability in luminosity but has always been observed in the soft `banana' branch of its color-color and HID \citep{Mondal_2020}. The binary inclination has been reported as low as \mbox{i $\le 10^{\circ}$}, using the variability and narrowness of optical emission lines \citep{Cornelisse_2013} and X-ray reflection modeling \citep{Miller_2013, Ludlam_2018,2025_Hall,2025_Ludlam_serx1}. However, some analyses of X-ray observations have reported a higher range of inclinations ($25^{\circ} < i < 50^{\circ}$) \citep{Cackett_2008, 2010_Cackett, Matranga_2017, chiang2016evolution, Mondal_2020}. It has a reported inner disk radius ranging from $\sim$ 7 -- 26 \rg, where \rg = $GM/c^2$ \citep{2010_Cackett,Miller_2013,chiang2016evolution,Chiang_2016_FeK,Matranga_2017,Ludlam_2018,2025_Hall}. The lower inferred radii and its correlation to luminosity strongly suggest the existence of BL between the edge of the disk and NS surface \citep{chiang2016evolution}. The large range of parameter values is a direct consequence of the lack of high energy resolution data from current X-ray instruments at the Fe K-alpha line $\sim6$~keV. For example, at 6 keV the spectral resolution of \nustar is 400 eV \citep{2013_Harrison_NUSTAR}, \nicer spectral resolution is 137 eV \citep{2016_Gendreau}, and the \chandra HETG has a spectral resolution of $\sim$30 eV at 6 keV \citep{2000_Weisskopf_CHANDRA}. The energy resolution of \xrism/Resolve \citep{XRISM_2018_Tashiro} provides the first opportunity to reveal the fine structure within the broadened reflection features. With an energy resolution of 5 eV at 6 keV, \xrism/resolve is the ideal tool for examining the broadening effects in the red- and blue-shifted wings of the Fe K line. The spectral data quality of Resolve now matches the level of that generated from reflection models and enables us to ascertain more robust system parameter values with better precision. Although initially designed for narrow line emission analysis, the Resolve instrument reveals how detailed reflection modeling of broadened lines can further advance our understanding of the physics and effects occurring in the region near compact objects in LXMBs.

\citet{2025_Ludlam_serx1} previously published on the \xrism/Resolve data alone, exploring different model prescriptions for the Fe line emission and investigating potential spectral differences between the upper banana, lower banana, and time averaged data. It was found that system parameters do not vary significantly between the areas of the HID and that relativistic reflection was the statistically best description of the observed spectrum. We expand on their analysis in this paper with the addition of simultaneous \nustar data of Ser X-1. Additionally, we test multiple continuum models on the joint dataset and analyze the lightcurves and spectra of the Type-I X-ray bursts that occurred during the observations. We test whether more physically motivated models can describe the continuum better than phenomenological models. This study represents the first analysis of simultaneous \nustar and \xrism data of Ser X-1. This paper also represents a comprehensive analysis of all available \xrism data on Ser X-1, examining persistent and burst emission with both \xrism instruments and \nustar. The combination of \nustar and \xrism data incorporates a wide passband with hard ($\ge10$ keV) and soft ($\le2$ keV) energy coverage, has high spectral resolution in the Fe-K region.

\begin{figure*}
    \centering
    \includegraphics[width=\textwidth]{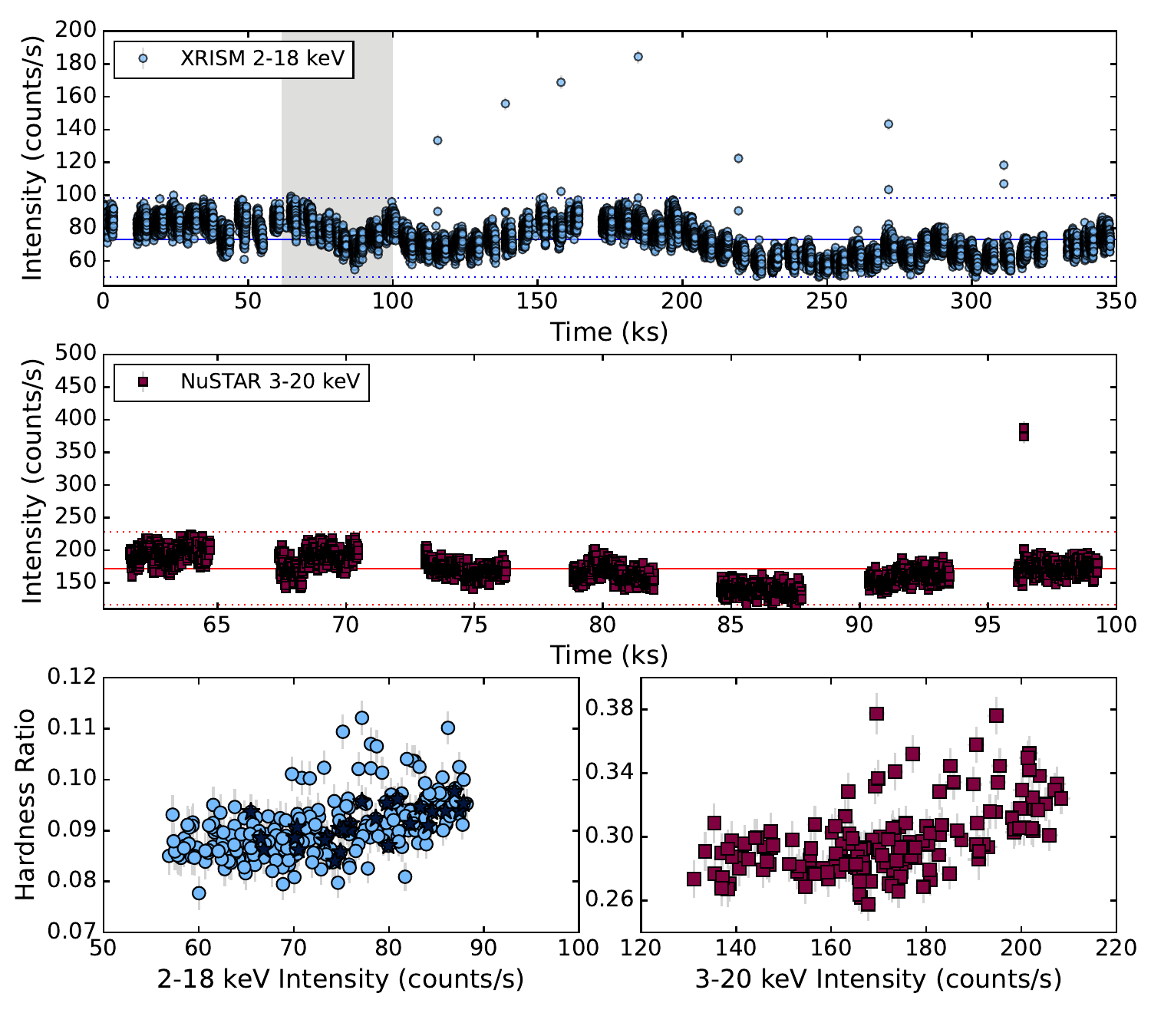}
    \caption{The light curves (using 10s bins) and HID for \xrism/Resolve (blue circles) and \nustar (red squares) observation of Ser X-1. The x-axis in the light curves represents time since the beginning of the \xrism observation. The Type-I X-ray bursts can be seen clearly as sharp increases of intensity compared to the persistent emission. The shaded region in the top panel indicates the time of \nustar observational overlap with that of \xrism. The stars in the bottom left panel represent the \xrism data during the overlap. The solid lines in the top two panels indicate the average flux level, while the dashed lines indicate the minimum and maximum persistent emission levels during the observations. The bottom panels are the 500s-bin HIDs with bursts removed, which show a consistent HR across the range of intensities for both \xrism and \nustar. We calculate HR for \xrism using the hard (10--16 keV) and soft (6--10 keV) bands, and likewise for \nustar, the hard (10--16 keV) and soft (6.4--10 keV) bands.}
    \label{XRNU_LC_HID}
\end{figure*}

\section{Observations and Data Reduction} \label{sec:data}
\subsection{XRISM}
\xrism observed Ser X-1 on October 17, 2024 for approximately 171 ks of cumulative exposure (ObsID: 201073010). Due to the Resolve gate valve being closed, no filter was necessary to reduce the illuminating flux of the source on the detector. To avoid pileup due to the brightness of Ser X-1, XTEND was operated in 1/8 windowed burst mode. There were a total of seven Type-I X-ray bursts present in the 1s binned \xrism light curve and they were removed to produce the time averaged spectra and light curves. For the burst removal, our GTIs began 10s prior to burst ignition and spanned roughly 100s to isolate the burst emission. We present a separate analysis of the bursts in Section ~\ref{T1XRB}.  We follow version 2.3 of the \xrism Quick Start Guide with the publicly available CALDB version index 20250315 to reduce the data. 

For Resolve, as per the recommendation of the calibration team, pixel 27 was removed from the data and only the highest resolution events (``Hp") were considered and extracted for light curves and spectra. Light curves were extracted in the 2--18 keV, 2--4 keV, 4--6 keV, 6--10 keV, and 10--16 keV energy bands to create the HID. The persistent count rate did not exceed 100 counts s$^{-1}$ in the 2--18 keV range, and as such we do not expect any event loss. We investigated and found there to be no overlap between the event loss GTIs and GTIs of the cleaned event files confirming there to be no event loss in our persistent GTIs. The \xrism data traces out both the upper and lower banana branches in the HID. We define our Hardness Ratio (HR) for \xrism/Resolve as the ratio between hard (10--16 keV) and soft (6--10 keV) intensities. Finally, we rebinned the spectra using \textit{ftgrouppha} with the optimal binning scheme \citep{Kaastra_2016} with no minimum counts per bin. 

We reduced the Xtend data according to the Quick Start Guide, first searching the detector images for anomalous pixels in the source and background region. Our search showed no evidence of anomalous pixels in either the source or background region. Our source and background regions extended the full width of the detector in 1/8 window mode (a box with approximate side length of 75 pixels). Ser X-1 is a bright source (average Xtend count rate: 246.5 \cps), however in 1/8 window mode it is below the 10\% pile-up limit of the Xtend CCDs \citep{2024_Yoneyama_xtend_pileup}. We tested various excision regions around the source and found that there were no spectral shape differences. We proceed therefore with the non excision region for our analysis. We removed Type-I X-ray bursts and produced spectrum using \textit{xselect}, in the same way as with Resolve. 

\subsection{NuSTAR}
\nustar observed Ser X-1 for an exposure 15.32 ks during the \xrism observation (ObsID: 91001342002). Data were reduced using nustardas v2.1.4 and CALDB v.20250729. Due to Ser X-1 being bright with an excess of 100 counts s$^{-1}$, we set statusexpr=``(STATUS==b0000xxx00xxxx000)\&\&(SHIELD==0)" in NUPIPELINE. A region of 100" radius centered on the source on DET0 and a background region of the same size but sufficiently far from the source on DET2 were used for spectral and light curve extraction. The 3--20 keV light curves were inspected for Type-I X-ray bursts. One was identified and removed for the time-averaged analysis. As can be seen in Figure \ref{XRNU_LC_HID}, the \nustar burst does not appear in the \xrism data (shaded region indicating overlap). This is due to the differences in visibility from mission orbit and the offset of observing windows. A separate analysis for the burst will be presented alongside those in the \xrism data. Similar to \xrism, the \nustar observation traced out the upper and lower banana in the HID. For \nustar, our HR is defined as the ratio of hard (10--16 keV) to soft (6.4--10 keV) intensity. GTIs were created in MET to extract spectra for the appropriate branches. The spectra were optimally binned using \textit{ftgrouppha} with a minimum of 25 counts per bin.

\section{Analysis and Results} \label{sec:results}
Figure \ref{XRNU_LC_HID} shows the \xrism and \nustar light curves and hardness ratios (HR) during the observation. The data shows some variation of intensity throughout the observation, while the HR remains largely constant in the banana branch. This is consistent with previous observations of the source \citep{Cackett_2008,Miller_2013,chiang2016evolution,Ludlam_2018,2024_Ursini,2025_Hall}. During the observation, the source exhibited several Type-I X-ray bursts (Figure \ref{XRNU_LC_HID}). There were a total of eight bursts recorded by \xrism and \nustar.

\subsection{Persistent Emission}
We first present the results and analysis of the persistent emission with the Type-I X-ray bursts filtered out. The source exhibited a hardness ratio consistent with the banana branch, both upper and lower, throughout the observation. \citet{2025_Ludlam_serx1} investigated the differences between upper, lower banana, and time averaged data and found that spectral parameters agree within the 90\% confidence intervals. Thus, we do not split the observation and present the time-averaged persistent spectrum. Spectral modeling is performed with {\sc xspec} \citep{1996_Arnaud_XSPEC} v12.15.1 using C-stat \citep{1979_Cash} with errors given at the 90\% confidence interval.

\subsubsection{Continuum Fitting}
\label{subs:cont}
We consider the \nustar data from 3--25 keV, as at higher energies the data become background dominated. For \xrism, we consider Xtend from 0.6--10 keV and Resolve from 3--8~keV. We chose these bounds to avoid spectral distortion for both Xtend and Resolve (see Appendix \ref{appendix:SUB}) and remain consistent with recent literature values \citep{2025_Mizumoto}. We also note that the effective area for Xtend drops significantly outside the given energy range (see figure 6.3 in Xtend Proposers' Observatory Guide\footnote{\href{https://heasarc.gsfc.nasa.gov/docs/xrism/proposals/POG/Xtend_SXI.html}{Xtend POG}}). We account for any calibration differences between \nustar and the \xrism instruments using {\sc crabcor} in {\sc xspec}. {\sc crabcor} is not a native model to {\sc xspec}, it is a multiplicative model component comprised of a normalization constant and an $E^{\Delta\Gamma}$ term originally introduced in \citet{2010_Steiner_et_al}. We fixed the normalization constant to 1 for the FPMA spectrum of \nustar and allowed to vary for the FPMB and \xrism Xtend/Resolve spectra to account for flux calibration differences. Xtend and Resolve do not share the same normalization constants or $\Delta\Gamma$ term. The $\Delta\Gamma$ term is set to zero for both \nustar spectra, and allowed to vary for Xtend and Resolve to account for inherent spectral slope differences between missions. 
We account for the neutral hydrogen column density along our line of sight with {\sc tbabs}. Abundances are set to {\sc wilms} \citep{2000_Wilms} and cross sections to {\sc vern} \citep{1996_Verner} within {\sc xspec}.

When fitting the continuum, we chose to use the double-thermal hybrid model consisting of a single temperature blackbody ({\sc bbody}) and multicolor disk blackbody ({\sc diskbb}). We chose to model the Comptonization component in two ways, with a simple power-law (Model 1) and with the {\sc thcomp} \citep{2020_Zdziarski} model in {\sc xspec}. For {\sc thcomp}, we expanded the energy range for model computation from 0.01 to 1000 keV in 1000 logarithmic bins. Model 1 is also known as the ``hybrid model" \citep{2007_Lin} for NS LMXBs. While the power-law model is additive, {\sc thcomp} is a convolution model that describes the Compton scattering of seed photons by a hot corona. These seed photons can originate from either the central blackbody or the disk. We convolve {\sc thcomp} with {\sc diskbb} (Model 2) and {\sc bbody} (Model 3), supplemented by the other non-convolved thermal component as an additive component. 

We calculate the Bayes information criterion (BIC $=$ C-stat$+$k log(n), k is the number of free parameters and n is the number of bins; \citealt{1978_Schwartz_BIC}) for each model. We report BIC of 9146.15 (Model 1), 8491.99 (Model 2), and 8481.31 (Model 3). Statistically, the fit improves in both models using {\sc thcomp} with Model 3 being slightly favored between the two.  Despite having improved statistics, the values obtained from {\sc thcomp} are highly unphysical and not in agreement with previous studies of the source that use a Comptonization model \citep{Bhattacharyya_2007,Mondal_2020}. The photon index we find is very hard ($\Gamma \leq1.38$), much lower than any previously reported values from Comptonization for the source \citep{Mondal_2020,2025_Hall}. \citet{2025_Ludlam_serx1} found a harder than typical photon index using \xrism data alone ($\Gamma\le1.46$), but our best fit model is below this. NS LMXB's have shown a wide range of potential $\Gamma$ values while in the soft state, but none as low as what is found here \citep{2024_Ursini,2025_Fei,2025_Thomas_gx17p2,2026_Iaria}. NS LMXBs have been shown to exhibit extremely low photon indices $\Gamma\sim1-1.5$ \citep{Parikh_2017}, but only in extremely hard states, in which Ser X-1 has never been observed.

Likewise, we find exceedingly low electron temperatures of $kT_e=2.65\pm0.04$ keV and $kT_e=2.60\pm0.03$ keV for Models 2 and 3, respectively.  Using archival RXTE data, \citet{2017_Burke} found that most NS LMXBs exhibit electron temperatures characteristic of 15-25 keV, nearly an order of magnitude larger than our best fit value. Additionally, we also find covering fractions that are too low (i.e., $\it{cov\_frac}$\ $\leq 0.20$) which indicates that the model prefers the original seed photons of the blackbody source.Because of this, we proceed with a phenomenological description of the continuum.

Most previous studies have elected to keep the neutral hydrogen column density $(N_H)$ fixed at a nominal value of either $4.0\times 10^{21}$ cm$^{-2}$ or $4.4\times 10^{21}$ cm$^{-2}$ (values from \citet{1990DickeyLockman, 2016_HI4PI}, respectively) for consistency or due to lack of spectral coverage in the sensitive energy ranges below 2 keV \citep{Miller_2013,Chiang_2016_FeK,Matranga_2017,Mondal_2020,2025_Ludlam_serx1}.  Model 1 gives a value of $7.0 \times 10^{21}$ cm$^{-2}$ which agrees well with \citet{Ludlam_2018,2025_Hall}. We find a disk temperature of $1.60\pm0.02$ keV and normalization of $42\pm2$ km$^2$/(D/10 kpc)$^2$ cos(i). We find a single blackbody temperature blackbody of $2.21^{+0.02}_{-0.03}$~keV. The simple power-law of $2.57^{+0.04}_{-0.05}$ is slightly harder than previously reported values in \citet{2010_Cackett,chiang2016evolution,Mondal_2020,2025_Hall}, but agrees with \citet{Bhattacharyya_2007,Ludlam_2018}.

For Models 2 and 3, we find a neutral hydrogen column density $(N_H)$ of $5.5 \times 10^{21}$ cm$^{-2}$, which is slightly higher than values from the nominal values of \citet{1990DickeyLockman, 2016_HI4PI}. \citet{Corrales_et_al_2016} notes that measured $N_H$ values from X-ray absorption will be at minimum 25\% higher than those found by \citet{1990DickeyLockman}. A disk temperature of roughly $0.95\pm0.01$ keV is found from both Models 2 and 3 with high normalizations of $261\pm12$ and $238\pm10$ km$^2$/(D/10 kpc)$^2$ cos(i) for model 2 and 3, respectively. We report blackbody temperatures of $1.45^{+0.01}_{-0.02}$ keV and $1.46^{+0.2}_{-0.1}$ keV, with normalizations of $4.0\pm0.2\ L_{39}/D_{10}$\footnote{Source luminosity ($L_{39}$) in units of $10^{39}$ erg/s and $D_{10}$ is the distance to the source in units of 10 kpc.} and $5.5\pm0.1\ L_{39}/D_{10}$.

\begin{figure}
    \centering
    \includegraphics[width=0.9\linewidth]{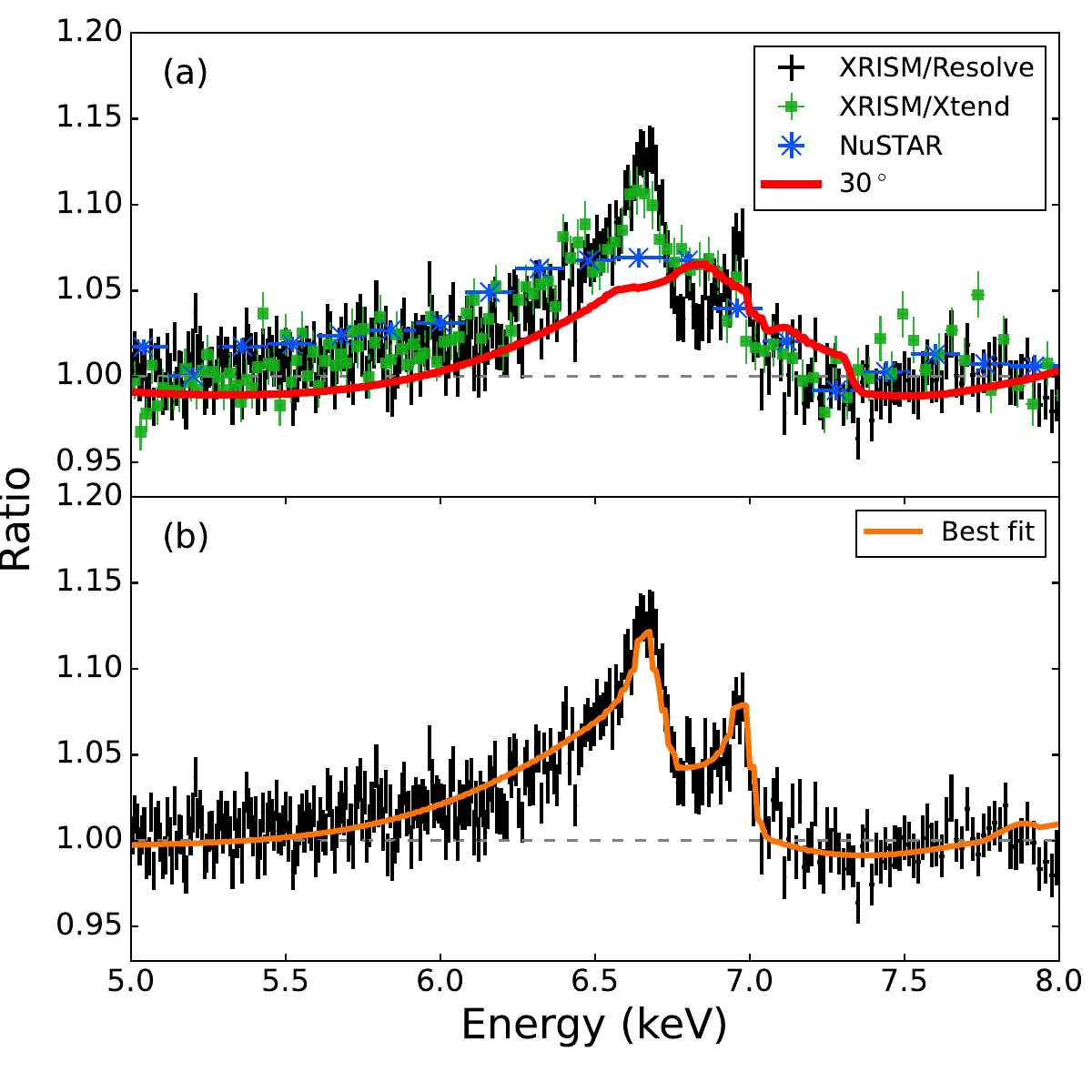}
    \caption{Panel (a) shows a ratio plot of Fe K region with \xrism/Resolve (black), \xrism/Xtend (green) and \nustar (blue). The red line represents the best fit reflection model when an inclination of $30^{\circ}$ is used with Resolve response files. Panel (b) shows the Resolve data with our best fit model overlaid.}
    \label{fig:Fe_line_Serx1}
\end{figure}

\input{Refl_table}\label{tab:refl_fit_xrnu}

\subsubsection{Reflection}
 We choose to model the reflection spectrum with a disk blackbody, reflection component, and a simple power-law for the high energy tail. This choice was similarly made in \citet{Ludlam_2018,2019_Ludlam_a,2025_Sudha_gx17p2,2025_Ludlam_serx1,2025_Hall}.
The shape of the Fe line in this source, as revealed by Resolve, is similar to that of GX 340+0 \citep{2025_Chakraborty_GX340p0,2025_Ludlam_GX340p0}. \citet{2025_Ludlam_serx1} explored alternative explanations for the observed line shape in the \xrism data of Ser X-1 and determined that relativistic reflection was required to fit the line. Thus, we proceed to model the \xrism and \nustar data sets simultaneously with the relativistic reflection model, \relxillns \citep{Garc_a_2022}. The model is a fully self consistent reflection model designed to describe the reflected emission from the accretion disk in NS systems. The model assumes a blackbody illumination spectrum with some characteristic temperature $kT_{bb}$. We tied our inner and outer emissivity indices together to act as a single emissivity index ($q$). Since the source is within our local galaxy, we set the redshift ($z$) at 0.  We fix the dimensionless spin parameter \mbox{($a=\frac{cJ}{GM^2}$)} at 0 given the fact that for the expected range of potential spins for Galactic NSs in LMXBs ($0\le a \le0.3$, \citealt{Galloway_2008,2011_Miller_Miller_Reynolds}) there is only a small resultant change in the location of the ISCO (6 \rg for $a=0$, 4.98 \rg for $a=0.3$).

The reflection fit is a significant improvement over the continuum fit of Model 1 (BIC = 3853.20, $\Delta$BIC = 4628.11) as well as providing a good description of the Fe line region. The full results are shown in Table \ref{tab:refl_fit_xrnu}. We find the disk to be fairly ionized with log($\xi$) $= 3.15$, with a low inclination ($i \le6^{\circ}$). This result matches well with previously reported values \citep{Ludlam_2018,2025_Hall,2025_Ludlam_serx1}. We would like to make particular note that with \xrism/Resolve, we can definitively determine the low inclination for Ser X-1, as evidenced by the steep blue wing present in both the Fe XXV and XXVI emission lines shown in Figure \ref{fig:Fe_line_Serx1}. We also show that when an inclination of 30$^{\circ}$ is assumed for the reflection model (i.e., near the values found in \citealt{2010_Cackett,Matranga_2017,Chiang_2016_FeK}), the model predicted line profile does not match the Resolve data. However, with the sensitivity of \nustar in the Fe K region, it is possible to see how larger inferred inclination measurements could arise.
As a check, we did perform reflection fits with {\sc thcomp}, replacing our power-law component as done with the continuum fitting. We do still find that the continuum parameters tend to unphysical values however, we find that all reflection parameters match within the 90\% confidence interval, signifying that our reflection results are not dependent upon the continuum model used.

\begin{figure}
 \centering
 \includegraphics[width=0.9\linewidth]{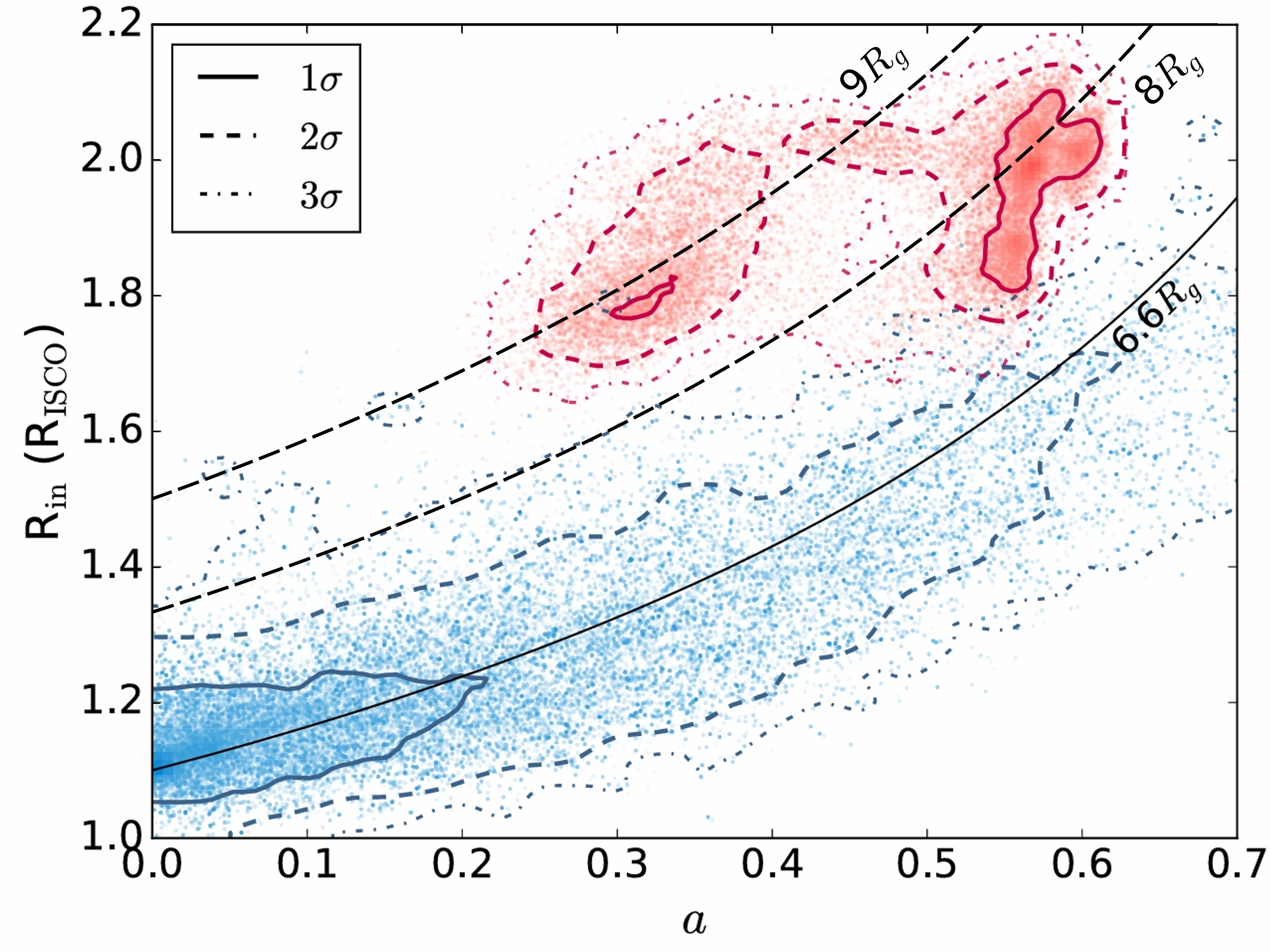}
 \caption{2D histogram of spin ($a$) vs inner disk radius (\rin). 1, 2, and 3$\sigma$ confidence intervals are shown for \xrism Resolve and Xtend (blue) and \nustar (pink). The \xrism data follows roughly a solid line corresponding to 6.6 \rg, signifying a unique radius for location of the inner disk edge. The \nustar data is largely insensitive to these parameters and follows no such trend, at points corresponding to two different unique radii (signified by the dashed lines at 8 \rg and 9 \rg).}
\label{fig:spin_Rin}
\end{figure}

 Since the spin of a compact object sets the location of the ISCO \citep{1972_Bardeen_Press_Teukolsky}, a degeneracy arises between spin and \rin. 
However, the location of the ISCO for a rapidly rotating compact object cannot be reproduced at lower spin values (i.e., a disk around a static compact object is unable to support stable circular orbits at radii smaller than 6~\rg). 
With sufficient S/N data, the relativistic effects imparted to the line profile may be resolved adequately to rule out a static NS (see \citealt{2023_Ludlam_HEXP} for simulations achievable with a future broadband X-ray mission). 
Historically, data quality has not allowed for an accurate determination of the dimensionless spin parameter of NSs in LMXBs through reflection.
We explore whether the increased spectral resolution available with Resolve combined with larger passband available through Xtend is sufficient to place constraints on the spin of Ser~X-1.

 As such, we performed additional fits with the \xrism and \nustar data separately allowing the spin parameter of \relxillns to be a free parameter that spans the entire range of possible values up to the breakup limit of 0.7 \citep{2025_Morales} with a soft limit of 0.3 (\citealt{Galloway_2008,2011_Miller_Miller_Reynolds}).
 We performed a Markov Chain Monte Carlo (MCMC) with 200 walkers, a burn length of $2 \times10^4$, and a chain length of $1\times10^5$. To highlight the difference in results that spectral resolution can give, Figure \ref{fig:spin_Rin} shows the results of the MCMC chain obtained using the data from the \xrism detectors and \nustar separate from one another. 
For \xrism, the confidence intervals for $a$ spanned the entire range of potential values and is still highly degenerate with the inner disk radius (see Figure \ref{fig:spin_Rin}). 
However,  for the first time, we do see that a unique value for the location of the inner disk is obtained as the data closely follows the value of $\sim$6.6 \rg regardless of the value for spin.
Conversely, the \nustar data does not select a single solution for the inner disk radius, but rather two solutions for \rin and spin; a consequence of the decreased spectral resolution of the detector.

\subsection{Type-I X-ray Bursts}\label{T1XRB}
\begin{figure*}
    \centering
    \includegraphics[width=\linewidth]{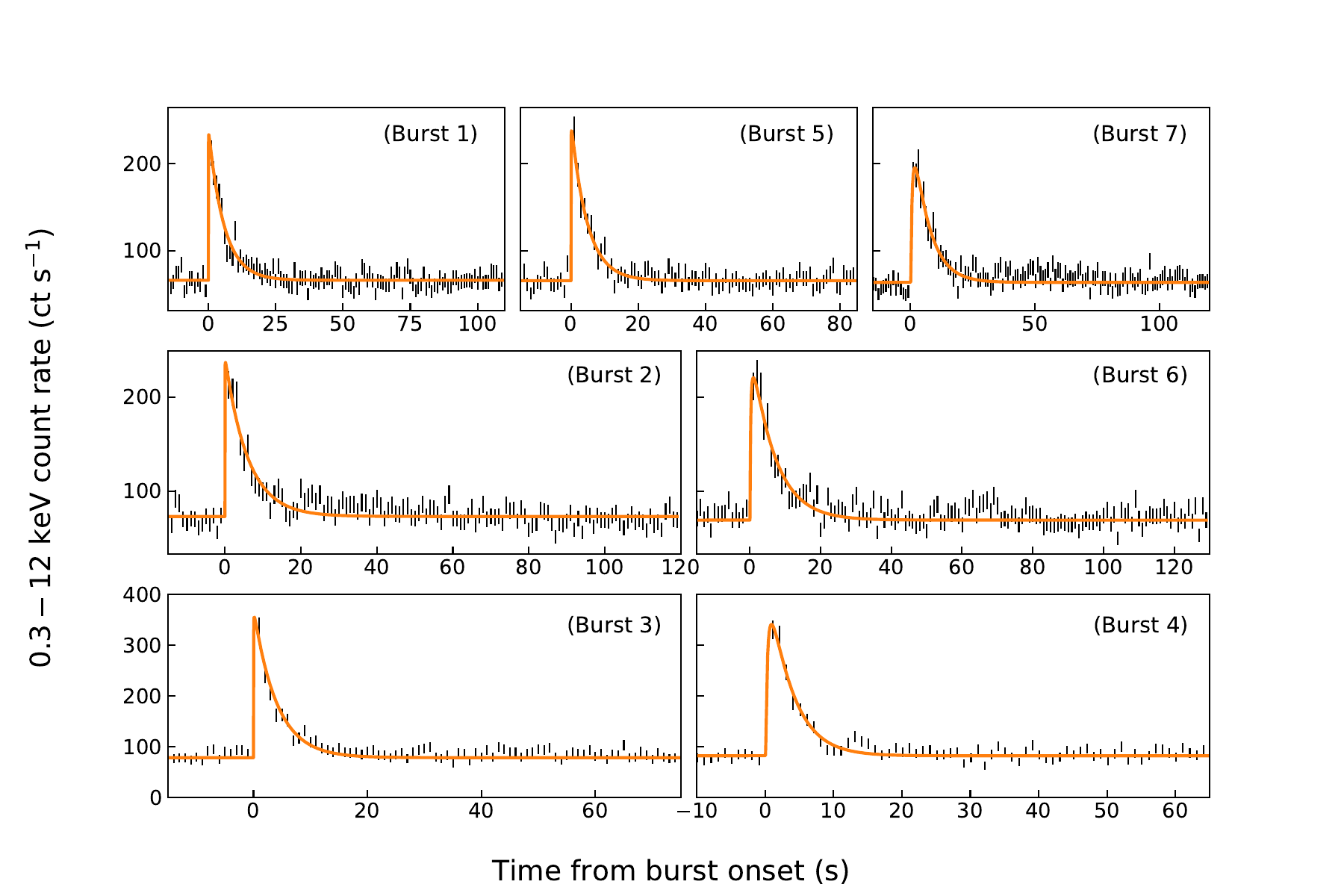}
    \caption{Grid of Type-I X-ray Bursts from \xrism in the 0.3 - 12 keV range. The rows correspond to the level of persistent emission (in cts s$^{-1}$) at the time of burst onset, low ($\le68.2$, row 1), mid ($68.2\le$ \cps $\le78.5$, row 2), high ($\ge78.5$, row 3) for clarity in presentation.} 
    \label{fig:xr_brst}
\end{figure*}

The source displayed 8 total Type-I X-ray bursts between \xrism and \nustar during the observation. The bursts last between 35--65 seconds, after which emission returns to persistent levels. None of the bursts showed evidence of photospheric radius expansion (PRE), nor were any of the bursts considered to be a superburst. We model the burst photometrically with a fast rise exponential decay (FRED) model \mbox{(Eq. 1).} 
\begin{equation}
f(t)= A\: \textrm{exp}\:[\frac{\tau_{R}}{t-t_{0}}-\frac{{t-t_{0}}}{\tau_{D}}]+C   \label{eq:FRED}
\end{equation} 
The full results can be seen in Table \ref{tab:burst_fit_xrnu}. The plot of the \xrism burst profiles and the best fit model for each burst is shown in Figure \ref{fig:xr_brst}. The rows are grouped by the persistent flux level at the burst onset. Despite the differing starting flux levels, we find no distinct correlation between blackbody temperature or inferred radius.

We treat the burst spectra in two ways, time averaged, taking the combined rise, peak, and decay spectra to search for potential reflection or absorption features and split into the aforementioned segments. As noted in Section \ref{sec:intro}, we modeled the bursts with a simple model of an absorbed blackbody on top of the underlying continuum, which we held fixed and treated as the background. This standard procedure uses the assumption that the underlying continuum emission does not change significantly during the burst, which has been found to be a good approximation \citep{1986_Van_Paradijs_Lewin,Galloway_2008}. 

We note that when screening for Hp events, we lose a significant number of events (i.e., our counts per second decreases to half the persistent emission levels due to the branching ratio during the high burst flux), and since the rise and peak times are roughly 1 -- 3 s in duration, we obtain very low signal-to-noise ratios (S/N $\lesssim5$). As a result, we include Mp events, using the command ``filter GRADE 0:1" when reducing in {\sc xselect}, to increase our S/N, as they have a similar energy resolution (7 eV) and our analysis of the segmented spectra is focused primarily on the broad blackbody shape.

We find and report the peak blackbody temperature throughout the burst segments, along with the associated normalization and inferred blackbody radius (the full set of parameter values can be seen in Table \ref{tab:spec_brst_xrnu}). We find an average blackbody radius of $\sim$10 km for all seven 
\xrism bursts and one \nustar burst. For the radius calculation, we assume spherically symmetric emission, a distance of 7.7 kpc, and a color correction factor of 1.7 \citep{Shimura_1995}.  
When searching for potential emission or absorption lines in the time averaged data we need the high resolution of the Hp and Mp events. Despite the increased number of events, we find that no burst spectra show visual evidence of statistically significant emission or absorption features. As a further test, we added gaussian absorption components to our model around $\sim6$ keV, where we would expect to see the features, leaving the centroid energy free to vary but none were conclusively identified. Although the combined datasets give us higher S/N for the bursts (avg $\sim55$), the data require a large amount of visual rebinning to show any shape whatsoever. Detailed spectral analysis of bursts will require the calibration of secondary and low resolution events. 

Additionally, to explore the existence and change in shape of a potential broad Fe line in the spectrum, we summed together larger 20s time bins of all bursts. We chose three 20 second intervals to explore, just prior to burst ignition, starting from the onset of burst ignition, and 20 seconds at the end of the tail as the flux returned to persistent levels. Unfortunately, we did not find evidence for a broad line in the Fe K region even when taking the summed 140s of emission with Hp and Mp events. Calibration of low resolution and secondary events may allow for this possibility in future analyses and observations of Type-I X-ray bursts in NS LMXBs.

\input{burst_table}
\input{burst_fred_table}

\section{Discussion} \label{sec:discussion}
We performed the first joint broadband spectral fits of simultaneous \xrism and \nustar data of the NS LMXB Ser X-1 to tightly constrain and characterize the reprocessed emission from the accretion disk in the 2 -- \mbox{25 keV} energy band.  The high spectral resolution afforded by \xrism/Resolve allows for a definitive exploration of the relativistic and geometric effects that shape the Fe K line. While \citet{2025_Ludlam_serx1} analyzed solely the \xrism/Resolve data set for Ser X-1, we include all available \xrism data from both Xtend and Resolve as well as the simultaneous \nustar observational data to present a comprehensive analysis of the source. We applied the self-consistent reflection model \relxillns to the data. We found tight constraints on the system inclination ($i=5.1^{\circ}\pm0.9^{\circ}$), and the location of the inner disk (to resolve the tension presented in current literature values (see Section \ref{sec:intro}). 

We explore constraining the spin of the central NS using the high spectral resolution of \xrism/Resolve in the Fe-K region.
 Reflection modeling has been shown to be an effective technique to determine the spins of accreting BHs (see \citealt{Reynolds2021} for a review) with many Galactic sources harboring near maximally spinning BHs ($a\geq0.9$: \citealt{Draghis2024}). At such high spin, emitted photons experience strong relativistic effects deep in the gravitational potential well as the ISCO approaches 1~\rg (see figures 1 and 2 in \citealt{Reynolds2021}), but this is a subtler effect for Galactic accreting NSs as the anticipated dimensionless spin parameter in these systems is significantly lower ($0\le a \le0.3$).
We find that although our data are insensitive to the spin of the NS, the location of the inner disk radius is well constrained to a unique value $\sim$6.6 \rg.  This result represents the first time a unique inner disk has been traced out in the spin-inner disk plane for a NS regardless of a spin measurement. The lack of tight constraints on spin are likely a result of insufficient S/N in our observation. Regardless, the power to constrain a unique radius with \xrism is very promising for future prospects on constraining NS spin.

If the estimated value of \rin and its contours in Figure~\ref{fig:spin_Rin} were to lie below the 6 \rg line, then a non-spinning NS could be ruled out. This is due to the fact that the inner disk radius of a static NS cannot extend below 6 \rg and therefore a smaller inferred radius cannot be replicated by a non-spinning NS (see \citet{2023_Ludlam_HEXP} and \citet{ludlam24} for further discussions of space-time metrics and implications for NS equation of state). Although a frequent burster, Ser X-1 has no current measurements of a burst oscillation frequency which can be used to infer the spin frequency of the NS \citep{2012_Watts}. If a burst oscillation frequency was known, it could offer a potential guide when reflection modeling. 

We tested three model prescriptions to describe the underlying continuum. Model 1 used a simple power-law to account for Comptonized emission from the corona alongside a multi-temperature disk blackbody and single temperature blackbody. Models 2 and 3 utilized the {\sc thcomp} model to describe Comptonization of thermal seed photons from either the disk (Model 2) or a central blackbody (Model 3) with the addition of whichever thermal component was not being Comptonized. For Model 1, our single temperature blackbody was found to have a temperature of $2.21^{+0.02}_{-0.03}$ keV (continuum) and $2.20\pm0.01$ keV (reflection), well in agreement with each other. These values are slightly lower than the values found in \citet{Miller_2013,chiang2016evolution,Chiang_2016_FeK,Mondal_2020} and \citet{2025_Hall}. We can attribute these differences to the degeneracy between the disk and blackbody temperatures and normalizations. Using the normalization of ($9.2\pm0.6$) $\times10^{-2} L_{39}/D_{10}$, we infer a blackbody radius of $12.2\pm0.7$ km. We use the distance of 7.7 kpc \citep{Galloway_2008} and a color correction factor of 1.7 \citep{Shimura_1995} similar to our calculations with the bursts in Section \ref{T1XRB}. This also agrees with the value obtained from the location of our inner disk (\rin = $13.6\pm1.2$ km for 1.4 \ms). We found a power-law index of $2.57^{+0.04}_{-0.05}$, notably lower than previously reported values in \citet{Miller_2013,Chiang_2016_FeK,chiang2016evolution,Mondal_2020,2025_Hall}. Overall, our best fit parameter values are consistent with other NS LMXBs in the soft state \citep{Ludlam20,2026_Das}. We find an overall BIC for Model 1 of 9146.15. Models 2 and 3 have lower BIC values (Model 2: 8491.99,  Model 3: 8481.31), and are therefore statistically favored; however, their parameter values are unphysical in some cases as we discussed in Section \ref{subs:cont}.

From our reflection modeling, we find an inner disk radius that extends nearly to the innermost stable circular orbit (ISCO) \rin = $1.1\pm0.1$\risco, consistent with the reports of other investigations \citep{Cackett_2008, 2010_Cackett, Miller_2013,Ludlam_2018, Matranga_2017, Bhattacharyya_2007, chiang2016evolution, Chiang_2016_FeK,2025_Hall,2025_Ludlam_serx1}. The slight truncation is largely believed to be due to the presence of a boundary between the edge of the disk and NS surface \citep{chiang2016evolution,2025_Hall,2025_Ludlam_serx1}. The spectral resolution of \xrism/Resolve revealed the dual-peaked structure within the Fe K line complex, showing both the Fe XXV and Fe XXVI peaks. 

We confirm the low inclination prescription of Ser X-1 as reported by \citet{2025_Ludlam_serx1}. We show that a higher inclination cannot replicate the line shape shown by the data. Figure \ref{fig:Fe_line_Serx1} demonstrates how the resolution of previous CCD based instruments could lead to higher inferred inclinations. The line shape produced at 30$^{\circ}$ shows much more similarity with the line shape from \nustar data than that of \xrism. The ionization parameter of the disk material was found to be $3.15\pm0.03$ which agrees well with \citet{Ludlam_2018} but is slightly higher than the values found in \citet{2024_Ursini} and \citet{2025_Hall}.  For the reflection fraction, we allow it to assume positive values to maintain the self-consistency between the illuminating and reflected emission and found a value of $0.19\pm0.01$, lower than value found by \citet{2025_Hall}, but higher than the value found in \citet{Ludlam_2018}. \citet{2024_Ursini} did use \relxillns to model joint \nicer, \nustar, and \ixpe data, however they do not report a value. We checked whether the results we obtained from reflection fitting may have depended upon the underlying continuum description and found that our reflection results agreed with the 90\% confidence level regardless of the continuum model used.

 We find that in our best fit model, the iron abundance is nearly 10$\times$ solar (the maximum value of the model). As the companion star is likely a $\sim1$ \ms star, this value is highly unlikely. This issue has been discussed at length in \citet{Garcia_2018_Fe_Abund,Tomsick_2018} and more recently \citet{2024_Ding} describes how updates to atomic data extending to higher densities and new \xstar routines can help potentially resolve this issue as it is believed to be a degeneracy with the density limit of the current publicly released model. Additionally, we performed reflection fits with the iron abundance set to  3$\times$ and 5$\times$ solar to investigate how it would affect our fitting results. In both cases, the fits worsen statistically when the abundance is fixed at a lower value. As we lower the iron abundance, we find that the inferred inner disk radius moves outward displaying an inverse correlation similar to the effect observed by \citet{Tomsick_2018} in the black hole LMXB Cygnus X-1. As a result, the emissivity index $q$ also tends to increasingly lower values with lower iron abundance, while the reflection fraction increases in strength to compensate.

We now compare our time-averaged continuum (Model 1) and reflection results directly to those in \citet{2025_Ludlam_serx1} to investigate whether the addition of Xtend and \nustar, and the extended passband they provide, impact our fitting results. With the addition on Xtend, we are able to probe the softer energies below the 1.7 keV cutoff of Resolve to get a value for our line of sight column density ($N_H$). Since only Resolve data was considered in \citet{2025_Ludlam_serx1}, they elected to fix the value at the nominal value of $4.4\times10^{21}$ cm$^{-2}$ \citep{2016_HI4PI}, while we find a higher value of $7\times10^{21}$ cm$^{-2}$. 
This difference alone will have a significant effect on the inferred values of the continuum as the absorption directly affects the shape of the spectrum at lower energies. For the disk and single temperature blackbodies, which are most affected, we find values of $1.60\pm0.02$ keV and $2.21^{+0.02}_{-0.03}$ keV, respectively. The values found in \citet{2025_Ludlam_serx1} are $0.68\pm0.01$ keV for the disk and $1.43\pm0.01$ keV for the single temperature blackbody. Our values are notably higher, however, we can similarly attribute this discrepancy to the degeneracy between the disk and blackbody temperatures and normalizations coupled with the differences in column density.

The addition of \nustar expands our passband to 25 keV, up from 17.4 keV with Resolve alone, to further constrain the Comptonized emission at higher energies. We find a powerlaw index of $2.57^{+0.04}_{-0.05}$, while \citet{2025_Ludlam_serx1} finds a best value of $1.7\pm0.1$ which is significantly harder. This discrepancy could be the result of calibration differences between the \xrism and \nustar instruments, we again note the limited Resolve passband in this work due to spectral shape differences above 8 keV (see discussion in Sec. \ref{appendix:SUB}). The differences in the thermal components can also affect the determination of the shape and strength of the Comptonized emission. Largely the discrepancies between the two results can be attributed to the limited passband of Resolve, as it lacks coverage to accurately determine the column density along the line of sight for softer X-rays ($<1.7$ keV) and lacks the hard X-ray ($\gtrsim17$ keV) coverage of \nustar.

Our reflection fitting results show some discrepancies with those found in \citet{2025_Ludlam_serx1}, however, the key system parameters of inclination and inner disk location match very well. Our best fit inclination of $5^{\circ}\pm1^{\circ}$ and inner disk radius of $\sim6$\rg are consistent with the values found in \citet{2025_Ludlam_serx1}. We believe that the inconsistency between the parameter values is again likely caused by the limited passband of Resolve alone. The lack of soft energy coverage of Resolve has a significant effect on the determination of many of the reflection parameters from \relxillns. In particular, our best fit value for the ionization parameter $log(\xi)=3.15\pm0.03$ is lower than the $log(\xi)=3.40\pm0.03$ value from \citet{2025_Ludlam_serx1}. Differences in the ionization parameter present themselves most strongly at energies below $\sim1$ keV, affecting the overall shape of the reflected spectrum as well as the abundance of emission and absorption lines \citep{2013_Garcia,2014_Garcia,Garc_a_2022,2024_Ding}. The soft energy coverage of Xtend allows us to better constrain the underlying thermal components and their relative strengths which informs the strength of the reflected spectrum as well. Our reflection fraction, directly related to the relative strength of the reflected spectrum, is found to be $0.19\pm0.01$ which is three times lower than the value found in \citet{2025_Ludlam_serx1}, again likely a consequence of the difference in spectral coverage in the softer energy ranges. However, we reiterate that key system parameters of inclination and location of the inner disk are recovered to similar values despite differences in other reflection fitting parameters and energy passbands.

During our observation, Ser X-1 exhibited eight total Type-I X-ray bursts. We performed photometric and spectral modeling for the bursts. We modeled the light curves with the FRED model. All burst profiles and parameters can be seen in Figure \ref{fig:xr_brst} and Table \ref{tab:burst_fit_xrnu}, respectively. Spectrally, we analyzed each burst to find the peak temperature during the burst, as well as in a time-averaged spectrum to search for any potential emission or absorption features. We modeled the bursts with an absorbed blackbody on top of the persistent emission which we treated as background. In the time averaged datasets, we found no evidence of any statistically significant features likely due to a lack of signal. It is possible that once the secondary and low resolution events are calibrated, it will allow for more detailed analysis with \xrism of burst spectra and an investigation of features present. Similar to \citet{2024_Ursini}, we investigated the burst phases to find the peak blackbody temperature of the burst emission. We find on average a peak temperature of $\sim2.1$ keV for the bursts and an average blackbody radius of $\sim10$ km. This temperature is largely consistent with other atoll sources that exhibit Type-I X-ray Bursts \citep{2022_Guver,2023_Thomas,2024_Yan,2026_Bhattacharya_bursts} and with the general trend of bursters which are seen to have peak temperatures between 2-3 keV \citep{Galloway_2008,Galloway_2021}. This agrees with our blackbody radius from continuum fitting as well as our expectation that the emission radius should be within the inner edge of the disk ($13.2$ km). Likewise, our results agree with \cite{2024_Ursini} within error bars. 

\section{Conclusion}
We performed spectral modeling on all available \xrism data using the Xtend and Resolve instruments on the NS LMXB Ser X-1, analyzing the persistent and burst emission. We included a simultaneous \nustar data set of the source as well to provide broad passband coverage up to 25 keV. We find that with the higher resolution \xrism data, while reflection modeling is not able to constrain a spin for the NS, the data are sensitive to a unique radius of the inner disk for the first time. This has implications for observations of other targets, for if a unique radius can be determined that lies at less than 6 \rg, it would rule out a static NS. Further calibration of all event types could also aid in the determination of NS spin in the future. We tested and found that the results of our reflection fitting did not depend upon the continuum model prescription used. Additionally, we performed the first spectral analysis of Type-I X-ray bursts for Ser X-1 with \xrism and find that the blackbody emission radius is $\sim$10 km, consistent with the location of the inner edge of the disk found from continuum and reflection modeling within errors. Due to the lack of calibration for secondary and low quality events, we lost a significant fraction ($\sim80\%$) of our source flux during the bursts when filtering. Improved calibration will enable detailed analysis of spectral shape, and aid in the search for potential absorption and emission features for other objects and future observations.\\

\textit{Acknowledgments:} This research has made use of data and/or software provided by the High Energy Astrophysics Science Archive Research Center (HEASARC), which is a service of the Astrophysics Science Division at NASA/GSFC. Support for this work was provided by the National Aeronautics and Space Administration (NASA) through grant No. 80NSSC25K7852.

\bibliographystyle{aasjournal}
\bibliography{references}

\newpage
\appendix{}
\counterwithin{figure}{section}

\section{Resolve Detector Subarray Regions}\label{appendix:SUB}
\begin{figure}
    \centering
    \includegraphics[width=0.7\linewidth]{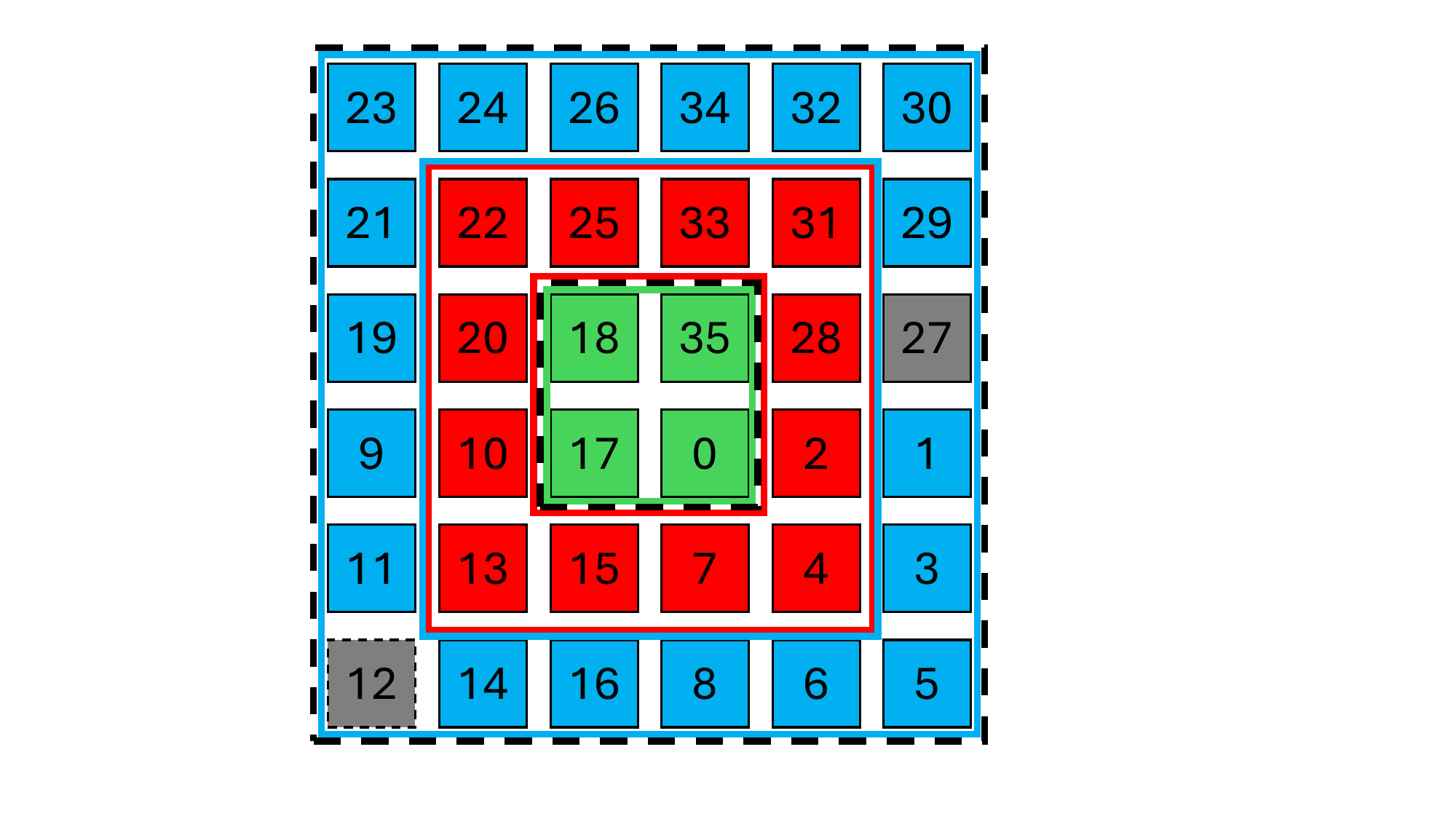}
    \caption{Schematic of the \xrism/Resolve detector with defined sub-arrays shaded and outlined. The regions are defined as central-four (green), inner ring (red), outer ring (blue), outer 32 (red + blue, dashed outline). Pixels numbers are shown, 12 (calibration) and 27 are greyed out as they were not included in analysis.}
    \label{fig:rsl_det_reg}
\end{figure}

Upon initial modeling of time averaged spectra, we discovered that the \nustar and \xrism/Resolve data differed in spectral shape above $\sim8$ keV. We note that the source was centered on the detector (pixels 0,17,18,35 in Figure \ref{fig:rsl_det_reg}). We divided the detector into sub-array regions to investigate whether or not different areas of the detector would produce differences in spectral shape. This process and potential spectral discrepancies were discussed in a presentation given at the \xrism workshop on Feb 3, 2025 at the University of Maryland \footnote{\href{https://heasarc.gsfc.nasa.gov/docs/xrism/analysis/workshops/doc_feb25/1_6_SDC.pdf}{XRISM Workshop}}. We split the detector into four distinct regions, the central four pixels, the outer 32 (30 when not including pixels 12 \& 27), an inner ring encircling the central four, and an outer ring encircling the inner ring which are indicated in Figure \ref{fig:rsl_det_reg}.

We briefly describe the data reduction process for the sub-array regions as they differ slightly to that described in the Quick Start Guide. We began by creating detector regions in \href{https://sites.google.com/cfa.harvard.edu/saoimageds9}{SAOImageDS9} from a detector image. In XSELECT we specified the desired pixels to be used from our cleaned event file via the command ``filter column ``PIXEL=0:0,17:18,35:35" " and extracted the events and spectra (this was for the central 4 pixels, a similar command was used for the other regions). For response file generation, we used the standard \textit{rslmkrmf}, \textit{xaexpmap}, and \textit{xaarfgen} commands with the specified region files and pixel specific event files created earlier. In order to effectively compare the impact of dectector regions, we chose the ``large" response file size.

Since the source was centered on the detector, we expect the central four pixels to have the highest count rates. Ser X-1 is a bright source, and as a result, when investigating the branching ratios, we found that the central four pixels showed the lowest ratio of Hp events when compared with other areas of the detector. However, due to the high count rates, the number of Hp events was still highest in the center. 
\begin{figure}
    \centering
    \includegraphics[width=0.7\linewidth]{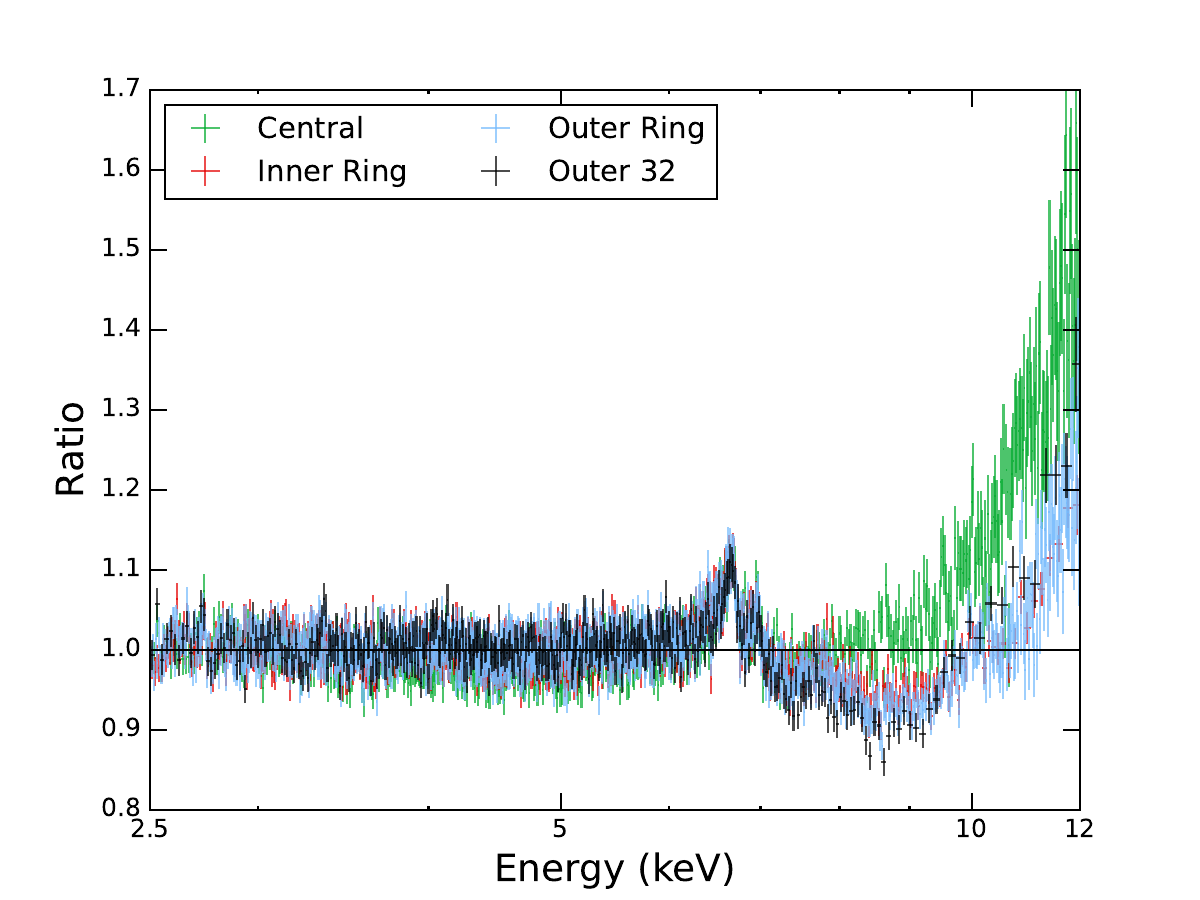}
    \caption{Fit residuals of different Resolve sub-array regions, highlighting the discrepancy in spectral shape above $\sim8$keV. The colors correspond to the detector regions shown in Figure \ref{fig:rsl_det_reg}. Note the large difference between the central pixels and the other 3 regions.}
    \label{fig:rslsub_reg_comp}
\end{figure}
When modeling the spectra we found that the outer and inner rings provide a very similar looking spectrum to their composite (the region excluding the central 4 pixels). Figure \ref{fig:rslsub_reg_comp} shows the resulting ratio plot when each region is fit with the same continuum model with the same model parameters. The ratio plots align well up until $\sim8$ keV, where a deviation can be seen with the central pixels showing a higher ratio compared with other regions. This indicates that there appears to be excess emission above 8 keV when using the central 4 pixels. No region matches well with the \nustar spectra above 8 keV (in terms of showing such excess in the ratio plot).

We consulted with the \xrism help desk and it was recommended that we restrict our modeling bandpass below $\sim$8 keV when using regions of more than one pixel \footnote{\href{https://xrism.isas.jaxa.jp/research/helpdesk/questions/0044.html}{XRISM Helpdesk Response}}. Although a limited bandpass (3--8 keV), the \xrism spectrum is still capable of highlighting how the exceptional spectral resolution, specifically in the Fe K-alpha region, enables us to tighten the constraints on the geometry of Ser X-1.

\end{document}

%% file: Refl_table.tex
\begin{table}[!t]
\begin{center}
\caption{Reflection Spectral Modeling of Simultaneous \xrism and \nustar Data}
\label{tab:refl_fit_xrnu}
\begin{tabular}{llcc}

\hline
Model                   & Parameter         & Values\\
\hline
\textsc{crabcor}    & $C_{FPMB}$ $(10^{-1})$ &$9.89\pm0.01$     \\
& $C_{Xtend}$ $(10^{-1})$ &$9.14\pm{0.08}$    \\
& $C_{Resolve}$ $(10^{-1})$ &$8.12\pm0.07$    \\
& $\Delta\Gamma_{Xtend}$ $(10^{-2})$ &$-0.83^{+0.6}_{-0.5}$  \\
& $\Delta\Gamma_{Resolve}$ $(10^{-2})$ &$5.5^{+0.5}_{-0.4}$  \\
\textsc{tbabs}       & $N_{H}(10^{21}\ \rm cm^{-2})$ &$5.91\pm0.03$          \\
\textsc{diskbb} & $kT$ (keV)          & $1.34\pm0.01$  \\
&norm       &$89\pm3$       \\
\textsc{powerlaw} & $\Gamma$         & $1.56\pm0.07$   \\
&norm  $(10^{-2})$          & $1.2\pm0.3$        \\
\textsc{relxillns} & $q$   &$2.23\pm0.03$       \\
&$i$ ($^{\circ}$) & $5.1\pm0.9$          \\
&$R_{in}$ (\risco)        & $1.1\pm0.1$         \\
&$R_{in}$ (\rg) & $6.6\pm0.6$                  \\
&$kT_{bb}$ (keV) &$2.20\pm0.01$ \\
&log $\xi$       &$3.15\pm0.03$ \\
&$A_{Fe}$               & $9.8^{+0.2}_{-0.3}$  \\
&$\log{n_{e}} (\rm cm^{-3})$    &$18.98^{+0.02}_{-0.04}$   \\
&\textit{$f_{refl}$} & $0.19\pm{0.01}$    \\
&norm$(10^{-3})$     & $3.62\pm0.04$                      \\

\hline
C-stat (dof)             & &3788.92 (2398) \\

\end{tabular}

\medskip 
\end{center}
Note.--- For \relxillns we fix the following parameters, break radius: $R_{br}$= 1000 $R_{g}$, spin: $a$ = 0, outer disk radius: $R_{out}$= 1000 $R_{g}$, redshift: $z$ = 0.
\end{table}

%% file: burst_table.tex
\begin{table*}[!t]
\begin{center}
\caption{Peak Blackbody Spectral Fits Type-1 X-ray Bursts XRISM/Resolve and NuSTAR}
\label{tab:spec_brst_xrnu}
\resizebox{\textwidth}{!}{
\begin{tabular}{l|ccccccc|c}

\hline
Parameter         &Burst 1 & Burst 2 & Burst 3 & Burst 4 & Burst 5 & Burst 6 & Burst 7 & NuSTAR Burst 1 \\
\hline
$kT$ (keV)          & $2.1^{+0.4}_{-0.2}$ & $2.1^{+0.2}_{-0.2}$ & $2.1^{+0.2}_{-0.2}$ & $2.2^{+0.1}_{-0.2}$ & $2.1^{+0.2}_{-0.2}$ & $2.2^{+0.2}_{-0.2}$ & $2.0^{+0.2}_{-0.2}$ & $2.4^{+0.1}_{-0.1}$ \\
norm ($10^{-2}$)      &$5.1^{+0.9}_{-0.3}$ & $4.9^{+1.0}_{-1.4}$ & $5.5^{+0.9}_{-0.7}$  & $6.1^{+0.7}_{-0.4}$  & $5.7^{+1.2}_{-0.6}$ & $5.3^{+0.4}_{-0.2}$  & $4.2^{+0.4}_{-0.2}$ & $7.95^{+0.2}_{-0.2}$ \\
Radius (km) & $10.1\pm3.4$ & $10.3\pm3.1$ & $10.2\pm2.4$ & $10.3\pm2.2$ & $10.3\pm4.1$ & $9.4\pm2.2$ & $10.1\pm2.9$ & $9.42\pm1.0$ \\
\hline
C-stat (d.o.f) &8803.8 (15997) & 8901.3 (15997) & 8728.6 (15997) & 8841.3 (15997) & 8230.8 (15997) & 8791.2 (15997) & 10493.8 (15997) & 1136.6 (1098) \\
\hline

\end{tabular}
}
\medskip 

Note.---Spectral fit of all Type-1 X-ray Bursts from XRISM/Resolve and NuSTAR using the simple model {\sc tbabs}$\times$({\sc bgd}+{\sc bbody}). The background ({\sc bgd}) is the persistent emission from Model 1.
 
\end{center}
\end{table*}

%% file: burst_fred_table.tex
\begin{table*}[!t]
\begin{center}
\caption{Lightcurve Modeling XRISM/Resolve $+$ NuSTAR FRED model}
\label{tab:burst_fit_xrnu}
\begin{tabular}{lccccccc|c}

\hline
 Parameter         &Burst 1 & Burst 2 & Burst 3 & Burst 4 & Burst 5 & Burst 6 & Burst 7 & NuSTAR Burst 1 \\
\hline
 $t_{0}$ (s) &115452   & 138851 & 158072 & 184759 & 219398 & 271263 & 311094 & 34810  \\
$\tau_{r}$ (s)          & 1.32 & 1.23& 1.51 & 1.26 & 2.53 & 1.16 & 2.50 & 3.8 \\
 $\tau_{d}$ (s)      & 57.1 & 59.8 & 40.2 & 32.8 & 47.6 & 64.6 & 61.6 & 51.3 \\
 $A$                 & 175 & 170 & 298 & 456& 184 & 207 & 233 & 663 \\
 $C$ (cts s$^{-1}$) & 66.0 & 72.9 & 78.4 & 82.0 & 65.5 & 69.1 & 63.5  & 167.7   \\

\hline
$\chi^2$/d.o.f  & 501.5/499 & 210.4/201 & 512.2/496 & 510.3/499 & 186.2/199 & 600.9/499 & 449.4/448 & 452.9/439 \\
\hline

\end{tabular}
\medskip 
\end{center}
Note.---Best fit parameters for Type-I X-ray bursts of Serpens X-1 using the Fast Rise Exponential Decay (FRED) model (Eq.~\ref{eq:FRED}).\\$t_{0}$- Burst onset time from start of observation\\$\tau_{r}$- Burst rise time\\$\tau_{d}$- Burst decay time\\$A$- Multiplicative constant\\$C$- Persistent counts/sec at burst onset.

\end{table*}